\documentclass{kluwer}    
\usepackage{graphicx}
\usepackage{epsfig}

\newdisplay{guess}{Conjecture}

\begin{document}                                                                                   
\begin{article}
\begin{opening}         
\title{On a Formation Scenario of Star Clusters}

\author{Christian \surname{Theis} \email{theis@astrophysik.uni-kiel.de}}  
\runningauthor{Ch.\ Theis}
\runningtitle{On a Formation Scenario of Star Clusters}
\institute{Institut f.\ Theoretische Physik und Astrophysik, Universit\"at Kiel,
24098 Kiel, Germany}
\date{November 28, 2001}

\begin{abstract}
     Most formation scenarios of globular clusters assume a molecular
  cloud as the progenitor of the stellar system. However, it is still 
  unclear, how this cloud is transformed into a star cluster, i.e.\ how
  the destructive processes related to gas removal or low star formation
  efficiency can be avoided. Here a scheme of supernova (SN) induced 
  cluster formation is studied. According to this scenario an 
  expanding SN shell accumulates the mass of the cloud.
  This is accompanied by fragmentation resulting in
  star formation in the shell. Provided the stellar shell expands 
  sufficiently slow, its self-gravity stops the expansion and
  the shell recollapses, by this forming a stellar system.

    I present N-body simulations of collapsing shells which move
  in a galactic potential on circular and elliptic orbits.  
  It is shown that typical shells ($10^5$ M$_\odot$, 30 pc)
  evolve to twin clusters over a large range of galactocentric distances.
  Outside this range single stellar systems are formed, whereas
  at small galactocentric distances the shells are tidally disrupted.
  In that case many small fragments formed during the collapse survive
  as single bound entities. About 1/3 of the twin cluster systems formed 
  on circular orbits merge within 400 Myr. On elliptic orbits 
  the merger rate reduces to less than 4\%. Thus, there could be
  a significant number of twin clusters even in our Galaxy,
  which, however, might be undetected as twins due to a large phase shift on
  their common orbit.

\end{abstract}

\end{opening}           


\section{Introduction}

  The exact formation process of globular clusters 
is still under debate. Suggested
mechanisms include -- among other scenarios -- e.g.\ the collapse of giant
molecular clouds (GMC) or the collision of molecular clouds
(e.g.\ \inlinecite{fall85}, \inlinecite{murray90}, \inlinecite{fujimoto97}). 
A common feature of most scenarios is the assumption of 
smooth initial gas distributions which are transformed into the cluster.
However, this assumption requires short formation timescales and unusually 
high star formation efficiencies in order to end up with a gravitationally 
bound system.
  An alternative model introduced by \inlinecite{brown91} can overcome these
difficulties: their scenario
starts with an OB-association exploding near the center 
of a molecular cloud. The expanding shell sweeps up the cloud material and in 
a later stage the expansion is decelerated and stopped by
the accumulated mass as well as 
the external pressure of the ambient interstellar 
medium. The shell itself is assumed to undergo fragmentation
and, finally, star formation. If these stars form a gravitationally bound
system, this stellar shell will recollapse, by this creating a
star cluster.
 
  At the moment a discrimination between different scenarios by  
direct simulations (starting from first principles) is far out of reach.
However, one can study different evolutionary stages in some detail.
E.g.\ \inlinecite{theis00} compared in a series
of N-body simulations the collapse of thin stellar shells
and homogeneous spheres in a galactic tidal field. These calculations 
were performed for circular and
eccentric orbits, but with a constant apogalacticon of 5 kpc. It was found that
collapsing shells preferably end in multiple systems, mainly twins, whereas
homogeneous spheres either form single clusters or become completely disrupted.

   In this paper the influence of the galactocentric distance, i.e.\ the
strength of the tidal field, on the collapse of stellar shells is 
investigated. Special forcus is put to the survival probability of the
formed multiple stellar cluster systems.

\begin{figure}[t]
  \centerline{\hbox{
  \psfig{figure=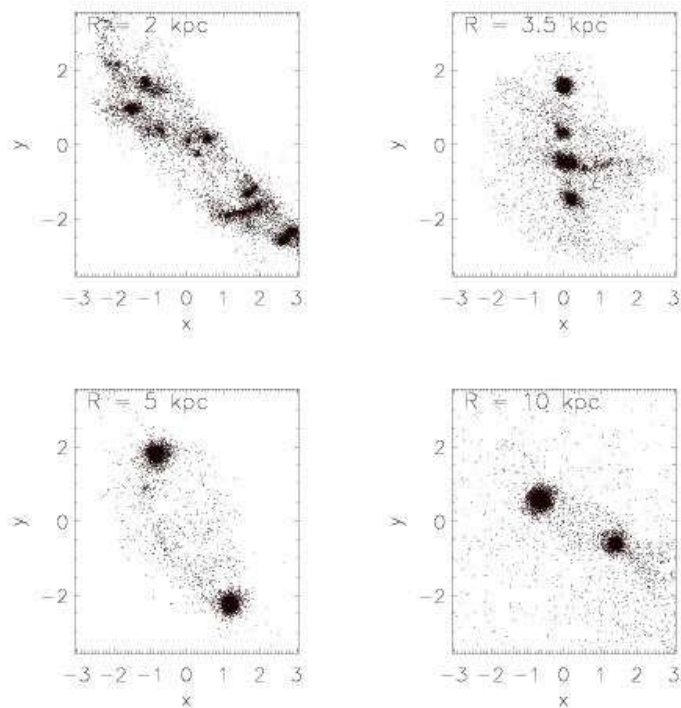,width=10.0cm,angle=0}
  }}
  \caption{Snapshots at $t=4\approx 30 \,\, \mbox{\rm Myr}$ for circular orbits 
     at different galactocentric distances:
     $R=2 \,\, \mbox{\rm kpc}$ (upper left),
     $R=3.5 \,\, \mbox{\rm kpc}$ (upper right),
     $R=5 \,\, \mbox{\rm kpc}$ (lower left),
     $R=10 \,\, \mbox{\rm kpc}$ (lower right). The spatial unit is 30 pc.  
   }
\end{figure}

\section{Numerical Models}

   The numerical models here start with a thin, spherical shell of
$10^5$ M$_\odot$, an outer radius of 30 pc and a thickness of 3 pc. 
The shell is initially at rest, i.e.\ there is no overall expansion or 
contraction of the shell with respect to its center. 
The potential of the Galaxy is modelled
by an isothermal halo with a circular speed of 220 km\,s$^{-1}$. The
investigated orbits correspond either to circular orbits or to an
elliptic orbit with an apo- to perigalacticon ratio of 10:1.
The calculations are performed with $N=10^4$ particles using a GRAPE3 board. 

   {\bf Circular Orbits.} Fig.\ 1 shows snapshots for collapsing shells
on circular orbits. The model starting at 5 kpc is typical for the models
resulting in a twin system. With its tidal radius of about 42 pc it is
stable against tidal disruption. However, the tidal field is strong
enough to delay the collapse along the direction to the galactic center.
By this, clumps form at the tips of this line ending up finally in the 
two clusters. At larger galactocentric distances the mass ratio of
both clusters increases: e.g.\ at 10 kpc two clusters are formed after
20 Myr which have a mass ratio of 5:2. The clusters formed here survive
until the end of the simulation at 400 Myr. 
20\% of the stars initially residing in the shell became unbound.
Beyond 11 kpc no twins, but single clusters are formed. 

  At 3.5 kpc the tidal radius is close to the initial radius of the shell.
However, the enhanced tidal field does not result in a disrupted system,
but in a less massive triple system accompanied by several smaller clusters.
The triple system dissolves quickly due to merging of two clusters. By
this, almost all stars of one cluster became unbound and a twin 
cluster system is left. 

  At about 2 kpc the tidal field prevents any collapse in the direction
to the galactic center. The fragments usually formed during the
collapse of a shell are then not destroyed in a violent collapse, but they
survive as gravitationally bound low-mass objects. E.g.\ 31 clumps
exist after $t=4$ and 12 of them survive the next 400 Myr. At the
end of the simulation 76\% of the stars are not bound to any cluster.

\begin{figure}
  \centerline{\vbox{
  \psfig{figure=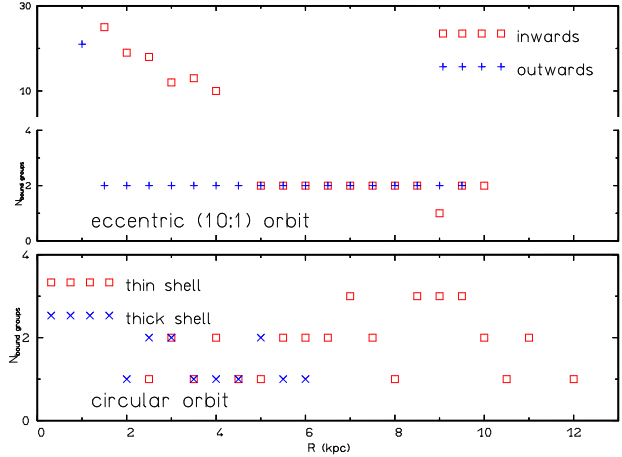,width=6cm,angle=270}
  }}
  \caption{Number of gravitationally bound clusters vs.\ initial galactocentric
   distance. Shown are results for $t=400 \,\, \mbox{\rm Myr}$ 
   (end of simulation)
   for circular (lower panel) and eccentric orbits (upper panel).
   The ''thick shell'' denotes simulations with an initial shell thickness
   of 15 pc. The other simulations are performed with a shell thickness of
   3 pc. ''Inwards'' and ''outwards'' corresponds to the initial phase 
   on the eccentric orbit.}
\end{figure}

 {\bf Survival Rates.}
  The simulations demonstrate that twin formation is expected over a large
radial range. On a longer timescale some twins are destroyed by merging,
e.g.\ for circular orbits about 1/3 of the twins merge within 400 Myr. 
The surviving twins are characterized by large
spatial separations which makes them less likely to undergo a subsequent
merger. Considering more realistic eccentric orbits, the merger rate
drops strongly: less than 4\% of the twins (i.e.\ one system!) undergoes
a merger. On the other hand, the fraction of disrupted systems increases 
to 20\%, because shells starting closer to perigalacticon can reach 
the ''disruptive zone'' in case of an eccentric orbit. However, the
survival probability for formed twins is not affected by this
destruction. Therefore, twin globulars might exist even in the 
Milky Way, but they could be unidentified as twins due to their large 
separation. Their characteristics (e.g.\ common orbit,
identical metallicity), however, might be used for an observational test 
of this cluster formation scenario.\\

{\bf Acknowledgements.}
The author is grateful to the organizers of the meeting for financial support.
The analysis of the cluster sizes has been performed with the SKID program 
kindly made available by the NASA HPCC ESS group at the University of 
Washington. The simulations were performed with the GRAPE3 
in Kiel (DFG Sp345/5).


\theendnotes

\end{article}

\begin{thebibliography}{}

\bibitem[\protect\citeauthoryear{Brown et al.}{1991}]{brown91}
Brown, J.H., Burkert, A., \& Truran, J.W. 1991, ApJ,  376, 115

\bibitem[\protect\citeauthoryear{Fall \& Rees}{1985}]{fall85}
Fall, S.M., Rees, M.J. 1985, ApJ, 298, 18

\bibitem[\protect\citeauthoryear{Fujimoto \& Kumai}{1997}]{fujimoto97}
Fujimoto, M., Kumai, Y. 1997, AJ, 113, 249

\bibitem[\protect\citeauthoryear{Murray \& Lin}{1990}]{murray90}
Murray, S.D., Lin, D.N.C. 1990, ApJ, 363, 50

\bibitem[\protect\citeauthoryear{Theis}{2000}]{theis00}
Theis, Ch. 2000, in Dynamics of Star Clusters and the Milky Way,
    S. Deiters et al.\ (eds.), in press (cf.\ astro-ph/0007144)
\end{thebibliography}
\end{document}